\def\bi{\bigskip}

\def\ct {\centerline}
\def\noi{\noindent}
\def\ii{\'{\i}}

\def\be{\begin{equation}}
\def\ee{\end{equation}}
\def\bea{\begin{eqnarray}}
\def\eea{\end{eqnarray}}
\def\al{\alpha}
\def\bet{\beta}

\def\nab{\nabla}
\def\pa{\partial}

\def\sqr#1#2{{\vcenter{\vbox{\hrule height.#2pt
        \hbox{\vrule width.#2pt height#1pt \kern#1pt
           \vrule width.#2pt}
        \hrule height.#2pt}}}}
\def\square{\mathchoice\sqr54\sqr54\sqr{6.1}3\sqr{1.5}6}
\documentstyle[12pt]{article}
\begin{document}
\baselineskip=25pt

\renewcommand{\thesection}{\Roman{section}}
\vspace*{.1in}
\begin{center}
  {\bf $\rm \Psi= W\, e^{\pm \Phi}$ quantum cosmological solutions for Class A
 Bianchi Models}
\normalsize

\vspace{18pt}
O. Obreg\'on and J. Socorro\\
\vspace{7pt}
{Instituto de F\ii sica de la Universidad de Guanajuato\\
P.O. Box E-143, 37150, Le\'on, Gto, M\'exico}\\
and\\
\vspace{7pt}
{Depto. de F\ii sica, Universidad Aut\'onoma Metropolitana-Iztapalapa\\
P.O. Box 55-534, 09340, Distrito Federal, M\'exico.}\\
\vspace{.5cm}
\end{center}
\begin{minipage}{5.3in}
{\abstract~~~~~
We find solutions for quantum Class A Bianchi models of the form
$\rm \Psi=W\, e^{\pm \Phi}$ generalizing the results obtained by Moncrief and
Ryan in standard quantum cosmology. For the II and IX Bianchi models there
are other solutions $\rm \tilde\Phi_2$, $\rm \tilde\Phi_9$ to the
Hamilton-Jacobi
equation for which $\rm \Psi$ is necessarely zero, in contrast with
solutions found in supersymmetric quantum cosmology.}

\end{minipage}

\vspace{15pt}

\vfill
\pagebreak

\renewcommand{\thesection}{\arabic{section}}
\renewcommand{\theequation}{\thesection.\arabic{equation}}

\section{Introduction}
\setcounter{equation}{0}

In recent years, progress has been made to find solutions{\cite {As,BrGaPu}}
to the canonical constraints of the full general relativity theory.
However, the  canonical
quantization program is far from complete. It is then hoped that the study
of some particular models could illustrate the behaviour of the general
theory. The Bianchi cosmologies are the prime example. Even in these
simplified cases little progress has been achieved. It
was just recently that solutions were found  for the more generic Bianchi
Class A models,  in particular the Bianchi IX model, resembling the situation
that one faces in the full theory.

It was first remarked by Kodama{\cite {Ko}} that solutions to the
Wheeler-DeWitt equation
(WDW) in the formulation of Arnowitt, Deser and Misner (ADM) and Ashtekar
(in the connection representation) are related by
$\rm \Psi= \Psi_A e^{\pm i \Phi_A}$, where $\rm \Phi_A$ is the homogeneous
specialization of the generating functional{\cite {As}} of the canonical
transformation from
the ADM variables to Ashtekar's. This function was calculated explicitely for
the diagonal Bianchi type IX model by Kodama{\cite {Ko}}, he also found
$\rm \Psi_A=const.$ as solution. Since $\rm \Phi_A$ is pure imaginary, for a
certain factor ordering, one expects a solution of the form
$\rm \Psi=W\, e^{\pm \Phi}$, W=const.,$\rm \Phi = i\Phi_A$. In fact
these type of solutions have
been found for the diagonal Bianchi type IX model{\cite {MoRy}}.
 For the special case of the Taub
model{\cite {MoRy,SeRy}} it is also possible to find a solution for which
$\rm W=const. \,e^{\al} \,e^{\bet_+}.$

In superquantum cosmology the same kind of solutions have been found by means
of two different approaches. Using supegravity N=1 it was shown{\cite{DeHaOb}}
for the
Bianchi type I model that the general solution has the form
$\rm \Psi= C_1 \, h^{-{1\over2}} \, e^{-\Phi} +
C_2 \, h^{-{1\over2}} \,\psi^6 \, e^{+\Phi}$, where $\rm C_1=C_2= const.$,
h is the determinant of the three metric, and $\rm \psi^{2n}$ express
symbolically the expansion of the wave function $\rm \Psi$ in even powers
(this guarantees Lorentz invariance) of the gravitino field.
The function $\rm \Phi$ in this particular case is zero but it was
suggested that for the Bianchi Class A models the solution has exactly this
form with their corresponding $\rm \Phi$ function. This conjeture has been
confirmed in a series of publications using the ADM
{\cite {De,AsTaYo}}
and the Ashtekar formulation {\cite {GuCa,CaOb}}. A more general postulate
for the Lorentz invariance seems to allow solutions also in the $\psi^2$ and
$\psi^4$ terms{\cite {Gr0}}. Similar solutions exist for a WDW equation
derived for the bosonic sector of the heterotic string{\cite {Li}}.

A second approach{\cite {Gr1}} considers the WDW equation also in the ADM
and Ashtekar formulation{\cite {ObPuRy}}, for the Bianchi model of interest and
proceeds by finding
appropriate operators which are the ``square root'' of this equation. This
procedure has the disadvantage that one has to introduce fermionic variables
without a direct physical meaning. However, for the physical quantities of
interest (like $\rm \Psi^* \Psi$) one integrate over these variables
{\cite {ObSoBe}}, getting
information about their influence on the unnormalized probability function.

The three previous procedures virtually result in the same kind of quantum
state and are of interest because for some of these models (by ex. Bianchi IX)
 these are the only known solutions. It is remarkable that they appear in the
three different approaches mentioned. However, these kind of solutions have
been found in standard quantum cosmology, only for the Bianchi type IX, Taub
and FRW models{\cite{MoRy,SeRy}}. The main point of this paper is to
generalize the results of Moncrief and Ryan to the diagonal Bianchi Class A
models. We will show that all solutions are  of the form
$\rm \Psi=W\, e^{\pm \Phi}$, where W is in general a function and can be
reduced to a constant for the Bianchi models VIII and IX, depending on the
factor ordering in the WDW equation.
For the Bianchi II and IX models there exist {\cite {GiPo}} others real
$\rm \tilde\Phi_2$ and $\rm \tilde\Phi_9$, however, it is surprising that
contrarely
to the results claimed in supersymmetric quantum cosmology{\cite {Gr0}}
$\rm \Psi=W\, e^{\pm \tilde\Phi_9}$ is not a solution of the WDW equation
because for this $\rm \tilde\Phi_9$, W is necessarely zero. Also W result to
be zero for $\rm \Psi=W\, e^{\pm \tilde\Phi_2}$.

Let us recall here the canonical formulation in the ADM formalism of the
diagonal Bianchi Class A models. The metrics have the form
\be
\rm ds^2= -dt^2 + e^{2\alpha(t)}\, (e^{2\beta(t)})_{ij}\, \omega^i \,
\omega^j, \label {met}
\ee
where $\alpha(t)$ is a scalar and $\rm \beta_{ij}(t)$ a 3x3 diagonal
matrix, $\rm \beta_{ij}= diag(x+ \sqrt{3} y,x- \sqrt{3} y, -2x)$,
$\rm \omega^i$ are one-forms that  characterize  each cosmological Bianchi
type model, and that obey
$\rm d\omega^i= {1\over 2} C^i_{jk} \omega^j \wedge \omega^k,$
$\rm C^i_{jk}$ the structure constants of the corresponding invariance
group.

\noi The ADM action has the form
\be
\rm I=\int (P_x dx+ P_y dy + P_{\alpha} d\alpha -  N {\cal H}_\perp) dt,
\label {acc}
\ee
where
\be
{\cal H}_\perp= \rm e^{-3\alpha}\Big(-P^2_\alpha+ P^2_x +P^2_y +
e^{4\alpha}V(x,y) \Big ),
\label {ham}
\ee
and $\rm e^{4\alpha}\, V(x,y)= U(q^\mu)$ is the potential term of the
cosmological model under consideration.

The WDW equation for these models is achived by replacing $\rm P_{q^\mu}$
by $\rm -i \partial_{q^\mu}$ in (\ref {ham}),
with $\rm q^\mu=(\al, x,  y)$. The factor $\rm e^{-3\al}$ may be
factor ordered
with $\rm \hat P_\alpha$ in many ways. Hartle and Hawking{\cite {HaHa}} have
suggested what might be called a semi-general factor ordering which in this
case would order $\rm e^{-3\al} \hat P^2_\al$ as
\be
\rm - e^{-(3- Q)\al}\, \pa_\al e^{-Q\al} \pa_\al=
- e^{-3\al}\, \pa^2_\al + Q\, e^{-3\al} \pa_\al,
\ee
where  Q is any real constant. With this factor ordering the Wheeler-DeWitt
equation becomes
\be
\rm \square \, \Psi + Q {\pa \Psi \over \pa \al} - U(q^\mu) \, \Psi =0,
\label {WDW}
\ee
where $\square$ is the  three dimensional d'Lambertian in the $\rm q^\mu$
coordinates.

The paper is then organized as follows. In Sec. II, we introduce the Ansatz
$\rm \Psi = W \, e^{-\Phi}$ in (\ref {WDW}) and set the general
equations for the Bianchi Class A models. In
Sec. III we present solutions for the cosmological Class A Bianchi models.
Only for the cases of Bianchi VIII and IX, W can be directly reduced to a
constant. For all other Bianchi models W is a function in contrast with the
solutions found in superquantum cosmology. For the Bianchi type
VI$_{\rm h=-1,}$  $\rm \Phi$ does not coincide with the general form that
appears in superquantum
cosmology. Sec. IV, is dedicated to final remarks.
\bi
\section{Transformation of the Wheeler-DeWitt equation}
\setcounter{equation}{0}

Under the Ansatz for the wave function
$\rm \Psi(q^\mu) = W(\al,x,y) e^{- \Phi},$
(\ref {WDW}) is transformed in
\be
 {\square \, W} - W {\square \, \Phi} - 2 {\nab W}\cdot {\nab \Phi} +
Q {{\pa W} \over {\pa \al}} -
Q W {{\pa \Phi} \over {\pa \al}} +
 W [ (\nab \Phi)^2 - U] = 0,
\label {mod}
\ee
where  $\rm \square = G^{\mu \nu}{\pa^2\over \pa q^\mu \pa q^\nu}$,
$\rm {\nab \, W}\cdot {\nab \, \Phi}=G^{\mu \nu}
{\pa W \over \pa q^\mu}{\pa \Phi \over \pa q^\nu}$,
$\rm (\nab)^2= -({\pa \over \pa \al})^2 +({\pa \over \pa x})^2 +
({\pa \over \pa y})^2$, with
$\rm G^{\mu \nu}= diag(-1,1,1)$,  U is the
potential term of the cosmological model under consideration.

\noi If one can solve the non-linear equation
\be
\rm (\nab \Phi)^2 - U = 0, \label {hj}
\ee
for $\Phi$, then one can obtain a master equation for the function W.
\be
\rm  {\square \, W} - W {\square \, \Phi} - 2 {\nab W}\cdot {\nab \Phi} +
Q {{\pa W} \over {\pa \al}} -
Q W {{\pa \Phi} \over {\pa \al}}  = 0.
\label {wdwmo}
\ee
(\ref {hj}) is  the classical
Einstein-Hamilton-Jacobi equation which can be obtained by replacing the
momentum
$\rm P_{q^\mu}\rightarrow {\pa \Phi \over \pa q^\mu}$ in
(\ref {ham}).

We were able to solve (\ref {hj}), for the Class A Bianchi models, by
doing
the following change of coordinates $\rm \beta_1=\al+x+\sqrt 3 y$,
$\rm \beta_2=\al+x-\sqrt 3 y$, $\rm \beta_3=\al-2x$, and rewrite
(\ref {hj})
in these new coordinates. With this change, the function $\Phi$ is obtained
and will be given in section III, in general form for the Class A
Bianchi Models.
In particular, Moncrief and Ryan{\cite {MoRy}}, have found in the case of the
Bianchi type
IX model an exact solution for  (\ref {hj}), being
\be
\rm \Phi= {1\over 6} e^{2\al}\lbrack e^{-4x}+ 2 e^{2x}\, cosh(2{\sqrt 3}y)
\rbrack,
\label {phi}
\ee
and then the solution for the wave function,
where  W=const., impliying Q= -6, and a solution for the Taub model
where the value of
the Q parameter is zero and $\rm W= const. \, e^{\al +x}$.

Let us make an assumption which will allow us to solve more easily
(\ref {wdwmo}), we demand that
\be
\rm {\square \, W} + Q{\pa W\over \pa \al}=0,
\label {wdwho}
\ee
which should be  consistent with
\be
\rm W {\square \, \Phi} + 2 {\nab \, W \cdot \nab \, \Phi} + QW
{\pa \Phi \over \pa \al} = 0,
\label {cons}
\ee
(\ref {wdwho}) is easier to solve than the original (\ref {WDW}),
because it does not contain any potential.

In the rest of this work, we will study the different solutions to Class A
Bianchi
models, where the Q parameter corresponds to different factor orderings in
the quantum
Wheeler-DeWitt equation.
\bi
\section{$\rm \Psi= W \, e^{\pm \Phi}$ solutions}
\setcounter{equation}{0}

In this section, we obtain the solutions to the equations that appear in
the decomposition of the WDW equation,
(\ref {hj}), (\ref {wdwho}) and (\ref {cons}) and give them for
the Class A Cosmological Bianchi models.

\noi Let us present, by means of a different procedure, the already known
solution{\cite {MoRy}} to (\ref {hj}) for the Bianchi type
IX model, because this procedure is used for the other Bianchi Class A
models.

Using the change of variables
$\rm (\al, x, y)\rightarrow (\bet_1, \bet_2, \bet_3)$, where the law of the
transformation between both set of variables is
\be
\begin{array}{ll}
\bet_1 = \al + x + \sqrt 3 y,\\
\bet_2 = \al + x - \sqrt 3 y,\\
\bet_3 = \al - 2x,
\end{array}
\label {trans}
\ee
the equation $\rm [\nab]^2= -({\pa\over \pa \al})^2 +({\pa\over \pa x})^2
+({\pa\over \pa y})^2$ can be written in the following way

\bea
\Big[\nab\Big]^2 &=& 3\Big[({{\pa}\over {\pa \bet_1}})^2+
({{\pa}\over {\pa \bet_2}})^2+
({{\pa}\over {\pa \bet_3}})^2 \Big ] - 6\Big[
 {{\pa}\over {\pa \bet_1}}{{\pa}\over {\pa \bet_2}} +
 {{\pa}\over {\pa \bet_1}}{{\pa}\over {\pa \bet_3}} +
 {{\pa}\over {\pa \bet_2}}{{\pa}\over {\pa \bet_3}}
\Big ]\nonumber\\
 &=& 3  \Big ( {{\pa}\over {\pa \bet_1}}+
                   {{\pa}\over {\pa \bet_2}}+
                   {{\pa}\over {\pa \bet_3}} \Big )^2 -12 \Big [
{{\pa}\over {\pa \bet_1}}{{\pa}\over {\pa \bet_2}} +
 {{\pa}\over {\pa \bet_1}}{{\pa}\over {\pa \bet_3}} +
 {{\pa}\over {\pa \bet_2}}{{\pa}\over {\pa \bet_3}}
\Big ].
\label {nabla}
\eea

The potencial term of the Bianchi type IX is transformed in the new variables
as
\be
U =  {1\over 3} \Big [ \Big ( e^{2\bet_1}+e^{2\bet_2}+e^{2\bet_3} \Big )^2-
4 e^{2(\bet_1+ \bet_2)} -4 e^{2(\bet_1+ \bet_3)}-4 e^{2(\bet_2+ \bet_3)}
\Big ].
\label {pot}
\ee
\noi Then (\ref {hj}) for this models is rewritten  in the new
variables as
\bea
 & & 3  \Big ( {{\pa \Phi}\over {\pa \bet_1}}+
              {{\pa \Phi}\over {\pa \bet_2}}+
              {{\pa \Phi}\over {\pa \bet_3}} \Big )^2 -12 \Big [
{{\pa \Phi}\over {\pa \bet_1}} {{\pa \Phi}\over
{\pa \bet_2}} +
{{\pa \Phi}\over {\pa \bet_1}} {{\pa \Phi}\over
 {\pa \bet_3}} +
{{\pa \Phi}\over {\pa \bet_2}} {{\pa \Phi}\over
{\pa \bet_3}}\Big ] \nonumber\\
& &\mbox{} -{1\over 3} \Big [ \Big ( e^{2\bet_1}+e^{2\bet_2}+e^{2\bet_3}
\Big )^2
-4 e^{2(\bet_1+ \bet_2)} -4 e^{2(\bet_1+ \bet_3)}-4 e^{2(\bet_2+ \bet_3)}
\Big ] =0.
\label {hanv}
\eea
Now, we can use the separation of variables method to get
solutions to the last equation for the $\Phi$ function, obtaining for the
Bianchi type IX model{\cite {MoRy}}
\be
\Phi_9= \pm {1\over 6} \Big ( e^{2\bet_1} + e^{2\bet_2} + e^{2\bet_3}
\Big ),
\label {phi9}
\ee
and {\cite {GiPo}}
\be
\tilde\Phi_9= \pm {1\over 6} \Big ( e^{2\bet_1} + e^{2\bet_2} + e^{2\bet_3}
-2\, e^{(\bet_1+\bet_2)}-2\, e^{(\bet_1+\bet_3)}-2\, e^{(\bet_2+\bet_3)}
\Big ).
\label {phi9n}
\ee
But surprisingly enough this $\rm \tilde\Phi_9$ does not produce any new
wave function because necessarely W=0. This means that the recent
solutions that have been claimed in supersymmetric quantum
cosmology{\cite{Gr0}} are not solutions of the standard WDW equation.

This same procedure is used for  getting the $\Phi$ function for the others
Bianchi Class A models. Also for the Bianchi type II model there exist a
second solution
$\rm \tilde\Phi_2=\pm {1\over 6}e^{2\bet_1} + F[(\bet_1+\bet_2)]$, where F is
any function of the argument. But for $\rm F\not= 0$ the wave function
vanishes (for F=0, $\tilde \Phi_2\equiv \Phi_2$). We
show these results in the table 1.

\noi With this result, the solution to (2.5) and (2.6), give for the W
function
\be
\rm  W_9= W_0 \, exp [(3+{Q\over 2})\al].
\label {w9}
\ee
where $\rm W_0=const,$ and $\rm Q=\pm 6$., then the wave function has the
following form
\be
\rm  \Psi_9= W_0 \, exp [(3+{Q\over 2})\al] \, exp [\pm \Phi_9].
\label {psi}
\ee

In the case of the Taub model, one replace in all terms only $\rm y=P_y=0$
\be
\rm \Phi_{TAUB}={1\over 6}\, e^{2\al}[2\, e^{2x} + e^{-4x}],
\label {taubpsi}
\ee
and the function W
\be
\rm W= W_0 exp (\al + x).
\label {taubw}
\ee

In this last case, the only value of the Q parameter is zero. These solutions
were
given by Moncrief and Ryan{\cite{MoRy}}.

\noi In the case of FRW model, the value of Q=2 and W=constant are obtained by
means of this method.
The solution is well known
$\rm \Phi_{FRW}= {1\over 2} e^{2\al},$ and
$\rm \Psi_{FRW}=W_0 exp[\pm \Phi_{FRW}].$
\bi

The functions W for the Bianchi Class A models are shown in Table 2.
\bi

 \bi
If one look at  the expressions for the functions $\Phi_i$, one notes that
there exist a general
form to write them using the 3x3  matrix $\rm m^{ij}$ that appear in
the
classification scheme of Ellis and MacCallum{\cite {RyShe,EllMac}}, the
structure constants are written in the form
\be
\rm C^i_{jk}=\epsilon_{jks} \, m^{si} + \delta^i_{[k} a_{j],}
\label {estr}
\ee
where $\rm a_i=0$ for the Class A models.

If we define $\rm g_i(q^\mu)= (e^{\bet_1}, e^{\bet_2}, e^{\bet_3}),$ with
$\rm \bet_i$ given in (\ref {trans}), all  solutions to (\ref {hj})
can be written as
\be
\rm \Phi(q^{\mu})= \pm {1\over 6}\, [g_i \, M^{ij} \, (g_j)^T],
\label {phig}
\ee
where $\rm M^{ij}=m^{ij}$ for the Class A Bianchi models, excepting the
Bianchi type $\rm VI_{h=-1}$ for which we redefine the matrix to be
consistent with (\ref {phig})
$$\rm  M^{ij} = \frac{6}{\sqrt 3} \,y \pmatrix {0 & 1 & 0 &\cr 1 & 0 & 0 &\cr
0 & 0 & 0}.$$
For the rest of the models (\ref {phig})  can be reduced to the expression
given previously in the literature, in connection with superquantum cosmology
{\cite {AsTaYo,Li}}
\be
\rm \Phi(q^{\mu})= \pm {1\over 6}\, [ m^{ij} \, g_{ij}],
\label {phign}
\ee
where $\rm g_{ij}$ is the 3-metric.
\bi
Then, for the Class A Bianchi models the wave function $\Psi$ can be written
in the general form
\be
\rm \Psi= W \, exp\,[\pm {1\over 6}\, [g_i \, M^{ij} \, (g_j)^T]],
\label {psig}
\ee
and for each cosmological model under consideration the wave function of
interest can be read from tables 1 and 2.
\bi
\section{Final remarks}
\setcounter{equation}{0}
Wave functions of the form $\rm \Psi= W\, e^{\pm \Phi}$ are the only
kind of solutions already found in supersymmetric quantum cosmology and also
for the WDW equation defined in the bosonic sector of the heterotic strings.
Also for the Bianchi type IX model these are the only known solutions in
standard quantum cosmology. We have found solutions of the same kind to the
Class A Bianchi models. It is to be noted that $\rm \Psi=W\, e^{\pm
\tilde\Phi_9}$
 (and $\rm \Psi=W\, e^{\pm \tilde\Phi_2}$ too) which seems to be a solution
{\cite {Gr0}} in supersymmetric quantum cosmology is not on allowed
wave function in standard quantum cosmology. For the Bianchi $\rm VI_{h=-1}$
model, (\ref {phig}) should be used instead of (\ref {phign}) to get the
right $\rm \Phi_6$ and consequently the corresponding wave function.
\bi

\ct  {ACKNOWLEDGMENTS}
\noi This work was supported in part by  CONACyT Grant 4862-E9406.

\newpage
\noi FIGURE CAPTION

Table 1. Potential  U and $\Phi$ function for the Class A Bianchi Models.

Table 2. W function and constraints between the constants in the solutions
for Class A Bianchi models.

\newpage

\begin{table}[h]
\begin{center}
\begin{tabular}{||l||l||l||r}  \hline \hline
& &\\
{\bf  Bianchi type} & {\bf Potential U} & ${\bf \Phi}$ \\
\hline \hline
 & &\\
I &   0 &  0 \\
\hline \hline
 & &\\
II &  $\rm \frac{1}{3} e^{4\bet_1} $ & $\rm \pm {1\over 6} e^{2\bet_1}$ \\
\hline \hline
 & &\\
$\rm VI_{h=-1}$&$\rm \frac{4}{3}e^{2(\bet_1+\bet_2)}$  &
$\rm \pm {1\over 6}[2 (\bet_1-\bet_2)\, e^{(\bet_1+\bet_2)}]$\\
\hline \hline
 & &\\
$\rm VII_{h=0}$ & $\rm \frac{1}{3}[e^{4\bet_1}+e^{4\bet_2}-
2e^{2(\bet_1+\bet_2)}]$
& $\rm \pm {1\over 6}[ e^{2\bet_1}+ e^{2\bet_2}]$ \\
\hline \hline
 & &\\
VIII & $\rm
\frac{1}{3}[e^{4\bet_1}+e^{4\bet_2}+e^{4\bet_3}-2e^{2(\bet_1+\bet_2)}+2e^{2(\bet_1+\bet_3)}+2e^{2(\bet_2+\bet_3)}]$
  &
$\rm \pm  {1\over 6}[ e^{2\bet_1}+ e^{2\bet_2} - e^{2\bet_3}]$ \\
\hline \hline
 & &\\
IX &  $\rm
\frac{1}{3}[e^{4\bet_1}+e^{4\bet_2}+e^{4\bet_3}-2e^{2(\bet_1+\bet_2)}-2e^{2(\bet_1+\bet_3)}-2e^{2(\bet_2+\bet_3)}]$ &
$\rm \pm {1\over 6}[ e^{2\bet_1}+ e^{2\bet_2} + e^{2\bet_3}]$ \\
\hline \hline
\end{tabular}
\end{center}

\label {tab1}
\end{table}
\newpage
\begin{table}[h]
\begin{center}
\begin{tabular}{||l||l||l||r}  \hline \hline
 & &\\
{\bf Bianchi type}&  {\bf W}&
 {\bf constraint} \\
\hline \hline
 & &\\
I &$\rm exp\,[\vec x\cdot \vec k]$ & $\rm a^2-aQ -(b^2+c^2)=0$ \\
\hline \hline
 & &\\
II &$\rm exp[(3+{Q\over 2}-a)\al + (b-a)x- {b\over \sqrt3}y]$
& $\rm 108-72a+24ab-16b^2-3Q^2=0$ \\
\hline \hline
 & &\\
$\rm VI_{h=-1}$& $\rm exp\,[\frac{1}{4}(\al+x)]$ &
$\rm Q=0$ \\
\hline \hline
 & &\\
 $\rm VII_{h=0}$ & $\rm exp\,[(3+{Q\over 2}-a)\al - ax]$
& $\rm 36-24a-Q^2=0$ \\
\hline \hline
 & &\\
VIII &  $\rm exp\,[(3+{Q\over 2})\al]$ & $\rm Q=\pm  6$\\
\hline \hline
 & &\\
IX &  $\rm exp\,[(3+{Q\over 2})\al]$ & $\rm Q=\pm  6$\\
 \hline \hline
\end{tabular}
\end{center}

\label {tab2}
\end{table}

\newpage

\begin{thebibliography}{99}
\bibitem {As}Ashtekar A 1986 {\it Phys. Rev. Lett.} {\bf 57} 2244; 1989
	{\it Phys. Rev. D} {\bf 36} 1587.
\bibitem {BrGaPu}Br\"ugmann B, Gambini R and Pullin J 1992
	{\it Phys. Rev. Lett.} {\bf 68} 431.
\bibitem {Ko} Kodama H 1988 {\it Prog. Theor. Phys.} {\bf 80} 1024;
	1990 {\it Phys. Rev. D} {\bf42} 2548.
\bibitem {MoRy} Moncrief V and  Ryan M P 1991 {\it Phys. Rev. D} {\bf 44}
	2375.
\bibitem {SeRy} Mart\ii nez S and  Ryan M P 1983 {\it Relativity, Cosmology,
	Topological Mass and Supergravity}, Proceedings of the Fourth Silarg
	Symposium, Caracas, Venezuela,  Edited by C Aragone (World
	Scientific, Singapore).
\bibitem {DeHaOb} D'Eath P D, Hawking S W and Obreg\'on O 1993
	{\it Phys. Lett. B} {\bf 300} 44.
\bibitem {De} D'Eath P D 1993 {\it Phys. Rev. D} {\bf 48} 713.
\bibitem {AsTaYo} Asano M, Tanimoto M and  Yoshino N 1993 {\it Phys. Lett.
	B} {\bf 314} 303.
\bibitem {GuCa} Capovilla R and  Guven J 1994 {\it Class. Quantum Grav.}
	{\bf 11} 1961.
\bibitem {CaOb} Capovilla R and Obreg\'on O 1994 {\it Phys. Rev. D}
	{\bf 49} 6562.
\bibitem {Gr0} Csord\'as A and Graham R {\it Preprint} gr-qc/9502004.
\bibitem {Li} Lidsey J E 1994 {\it Phys. Rev. D} {\bf 49} R599; {\it Preprint}
	 FERMILAB-Pub-94-062-A.
\bibitem {Gr1} Graham R 1991 {\it Phys. Rev. Lett.} {\bf 67} 1381;
	 1993 {\it Phys. Rev. D} {\bf 48} 1602.
\bibitem {ObPuRy} Obreg\'on O, Pullin J and Ryan M P 1993 {\it Phys. Rev. D}
	{\bf 48} 5642.
\bibitem {ObSoBe} Obreg\'on O,  Socorro J and Ben\ii tez J 1993
	 {\it Phys. Rev. D} {\bf 47} 4471;
	 1994 {\it Proceedings of the  8th Silarg
	Symposium}, \'Aguas de Lind\'oia, Brasil Edited by P S Letelier
	and W A Rodrigues Jr (World Scientific, Singapore) 446.
\bibitem {GiPo} Gibbons G W and Pope C N 1979 {\it Commun. Math. Phys.}
	{\bf 66} 267.
\bibitem {HaHa} Hartle J and Hawking S W 1983 {\it Phys. Rev. D} {\bf 28}
	2960.
\bibitem {RyShe}  Ryan M P and  Shepley L C 1975 {\it Homogeneous
	Relativistic Cosmologies} (Princenton).
\bibitem {EllMac} Ellis G F R and MacCallum M A H 1969 {\it Comm. Math. Phys.}
	{\bf 12} 108.
%
\end{thebibliography}
\end{document}